# TWO-PHASE MULTI-DOSE-LEVEL PET IMAGE RECONSTRUCTION WITH DOSE LEVEL AWARENESS

*Yuchen Fei[1], Yanmei Luo[1], Yan Wang[1, *], Jiaqi Cui[1], Yuanyuan Xu[1], Jiliu Zhou[1], Dinggang Shen[2, 3]*

[1]School of Computer Science, Sichuan University, Chengdu, China
[2]School of Biomedical Engineering & State Key Laboratory of Advanced Medical Materials and Devices, ShanghaiTech University, Shanghai 201210, China
[3]Shanghai United Imaging Intelligence Co., Ltd., Shanghai, China

## ABSTRACT

To obtain high-quality positron emission tomography (PET) while minimizing radiation exposure, a range of methods have been designed to reconstruct standard-dose PET (SPET) from corresponding low-dose PET (LPET) images. However, most current methods merely learn the mapping between single-dose-level LPET and SPET images, but omit the dose disparity of LPET images in clinical scenarios. In this paper, to reconstruct high-quality SPET images from multi-dose-level LPET images, we design a novel two-phase multi-dose-level PET reconstruction algorithm with dose level awareness, containing a pre-training phase and a SPET prediction phase. Specifically, the pre-training phase is devised to explore both fine-grained discriminative features and effective semantic representation. The SPET prediction phase adopts a coarse prediction network utilizing pre-learned dose level prior to generate preliminary result, and a refinement network to precisely preserve the details. Experiments on MICCAI 2022 Ultra-low Dose PET Imaging Challenge Dataset have demonstrated the superiority of our method.

## 1. INTRODUCTION

As a non-invasive nuclear imaging technique, Positron emission tomography (PET) plays a pivotal role in early disease diagnosis and intervention. In clinic, obtaining high-quality PET images necessitates the injection of standard-dose radioactive tracers into human body. Yet, the inherent radiation of the injected tracers inevitably raises potential health hazard [1]. Yet, reducing the tracer dose will involve unintended noises and artifacts, leading to inferior image quality. Thus, reconstructing high-quality standard-dose PET (SPET) from low-dose PET (LPET) is a promising alternative.

In recent years, with the rise of deep learning, a range of studies have been proposed to reconstruct high-quality SPET images from LPET images [2-5]. Particularly, Xiang *et al.* [2] designed a convolutional neural network (CNN) based model to gradually enhance the LPET image quality following auto-context strategy. Wang *et al.* put forward a 3D conditional generative adversarial network (GAN) [3] to obtain high-quality SPET images at low dose. Lei *et al.* [4] adopted the cycle consistency to boost the reconstruction performance of whole-body PET. Luo *et al.* [5] combined spectrum constrain and adaptive rectification for high-quality SPET synthesis.

Although existing methods have attained relatively promising results, they focus exclusively on learning the mapping relationship between LPET at a single dose level and their corresponding SPET. However, LPET images in clinic have the property of various dose reduction factors (DRFs) and correspondingly present different noise levels, leading to the undesired performance of current single-dose-level LPET image reconstruction methods when applied in versatile clinical scenarios. To the best of our knowledge, there are only a limited number of works turned to investigate the more challenging multi-dose-level LPET reconstruction problem. Specifically, Xue *et al.* [6] developed a modified conditional GAN model which was trained with LPET and SPET image pairs of all doses, to enhance multi-dose-level LPET image quality. Zhang *et al.* [7] integrated anatomical structural information from MRI images to boost the blind denoising performance of LPET images. The aforementioned multi-dose-level LPET reconstruction methods simply mix the LPET images of different dose levels for training. Yet, investigating the correlation between different images and their corresponding noise levels has been demonstrated to be conducive to constructing a robust denoising model [8]. In addition, these works ignore the distinctive characteristics of LPET images at each dose level, prohibiting the model from learning semantic representation effectively.

In this paper, motivated to address the above key issues, we propose a novel two-phase coarse-to-fine framework with dose level awareness for reconstructing high-quality SPET images from multi-dose-level LPET images, including a pre-training phase and a SPET prediction phase. Specifically, the pre-training phase introduces a tracer dose classification task to fully exploit the fine-grained discriminative features of LPET images and a self-reconstruction task for learning effective semantic representation, thus enhancing the feature extraction capability of the network. The SPET prediction phase adopts a coarse-to-fine design, i.e., the coarse prediction network (CPNet) initialized with pre-trained parameters generates a coarse prediction, and the refinement network (RefineNet) samples the residual between the coarse prediction and the target SPET image. By combining the coarse prediction and the residual, more realistic reconstructed PET (RPET) images can be finally obtained.

## 2. METHODOLOGY

The framework of our model is shown in Fig. 1, consisting of a pre-training phase and a SPET prediction phase. The pre-training phase employs two tasks, i.e., dose classification and self-reconstruction. The dose classification task aims to exploit the fine-grained discriminative features and the self-reconstruction task aims to learn the semantic representations of PET images, respectively. The SPET prediction phase takes a coarse-to-fine design, including a CPNet for generating a coarse prediction and a RefineNet to estimate the residual map between the coarse prediction and the target SPET. By incorporating both the coarse prediction and residual, our framework can effectively reconstruct high-quality RPET from LPET. The detailed architecture of our model will be described below.

### 2.1. Architecture

As shown in Fig. 1 (a), the PretrainNet is designed with reference to U-Net [9], taking LPET of varying DRFs as input and outputting self-reconstructed LPET through an encoder and a symmetric decoder. Specifically, the encoder contains four down-sampling blocks structured as 4×4 Convolution-BatchNorm-LeakyReLu. Similarly, the decoder adopts four up-sampling blocks with 4×4 Deconvolution-BatchNorm-ReLu, gradually restoring the features extracted by the encoder to the self-reconstructed LPET. Following [10], the skip connections are omitted to enforce the decoder to learn the mapping between features instead of directly copying the low-dimensional features from the encoder. A residual block which contains two 3×3 convolution and ReLU function, is embedded before each down-sampling block and up-sampling block to mitigate gradient vanishing and promotes the network convergence. Moreover, the dose classifier is adopted to explore fine-grained discriminative features of multi-dose-level LPET images. Concretely, the features extracted by the last layer of the encoder is flattened and the passed through a fully connected layer to categorize the input LPET into certain DRF. The mean squared error (MSE) loss and cross entropy (CE) loss are appliedto supervise accurate self-reconstruction and tracer dose classification, respectively. The overall loss can be defined as:

$$L_{pretrain} = \lambda L_{mse}(l, \hat{l}) + (1 - \lambda)L_{ce}(y, \hat{y}) \quad (1)$$

$$L_{mse}(l, \hat{l}) = \frac{1}{n}\sum_{i=1}^{n}(l_i, \hat{l}_i)^2 \quad (2)$$

$$L_{ce}(y, \hat{y}) = -\sum_{k=1}^{m} y_k \log(\hat{y}_k) \quad (3)$$

where $n$ is the number of samples, $y_k$ and $\hat{y}_k$ represent the real DRF label of $l$ belonging to the DRF category $k$ and the probability value that the model predicts the DRF of this sample to be of category $k$, respectively. And $m$ denotes the number of DRF categories which equals to 3 (DRF =20, 50, 100). $\lambda$ is the hyperparameter to balance the two loss terms.

### 2.2. SPET Prediction Phase

Acknowledging the reconstruction difficulty induced by the substantial variations in noise and artifact among LPET images of various dose levels, we introduce a coarse-to-fine strategy based on residual learning in the SPET prediction phase, as illustrated in Fig. 1 (b). The CPNet adopts the same structure as PretrainNet, aiming to predict a coarse-denoised PET image from the LPET image. The RefineNet employs an encoder-decoder structure with shallower layers, taking both the coarse prediction and LPET as input to estimate the residual between the coarse prediction and the SPET. Note that, the parameters of the CPNet encoder are initialized with those of the pre-trained PretrainNet for efficiently leveraging the dose-aware information and achieving superior feature extraction performance. By combining the coarse prediction and the residual, we can obtain more realistic RPET images. L1 loss is adopted to supervise the training of CPNet and RefineNet, as shown below:

$$L_{CPNet}(s, \tilde{s}) = \frac{1}{n}\sum_{i=1}^{n}|s_i - \tilde{s}_i| \quad (4)$$

$$L_{RefineNet}(r, \tilde{r}) = \frac{1}{n}\sum_{i=1}^{n}|r_i - \tilde{r}_i| \quad (5)$$

where $s$ and $\tilde{s}$ represent the coarse prediction and the target SPET image, respectively. We denote $r$ as the real residual between $\hat{s}$ and $s$, and $\hat{r}$ is the predicted residual. Thus, the total loss function of SPET prediction phase is formulated as:

$$L_{img\_pre} = L_{CPNet}(s, \tilde{s}) + \beta L_{RefineNet}(r, \tilde{r}) \quad (6)$$

where $\beta$ is the hyperparameter to balance the loss terms.

### 2.3. Implementation Details

All experiments are conducted on the PyTorch framework with four NVIDIA GeForce GTX 2070SUPER GPUs. In the pre-training phase, λ is initialized as 0 and then linearly increases from 0 to 1 as the epoch increases. In this way, the PretrainNet could prioritize the extraction of discriminative features in the initial training period by boosting the classification performance, and fully exploit the semantic information of input LPET in the later training by primarily focusing on improving the reconstruction performance. Thus, the PretrainNet could be comprehensively trained. During the SPET prediction phase, the CPNet is initialized with the parameters of the pre-trained encoder and is trained in an end-to-end manner with the RefineNet. The whole training process lasts for 100 epochs, using Adam optimizer with a learning rate 0.0002 and batch size 16. $\beta$ is empirically t to 1.

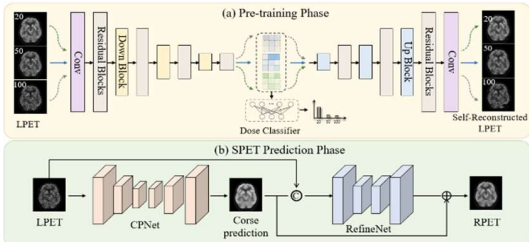

Fig. 1 Overall architecture of the proposed framework.

## 3. EXPERIMENTS AND RESULTS

We evaluate our method on a public brain dataset, which is obtained from MICCAI 2022 Ultra Low Dose PET Imaging Challenge Dataset [6]. It includes 206 subjects acquired by Siemens Biograph Vision Quadra System. 170 subjects are randomly selected for training, and the rest 36 for testing. For each subject, LPET images at six DRFs (i.e., 2, 4, 10, 20, 50, 100) and the corresponding SPET images are provided. In this study, only LPET images at DRFs of 20, 50, and 100 are employed. The PET image has a resolution of 128 × 128 × 128, and 2D slices along the z-coordinate are adopted for training and testing. Three standard metrics are utilized to quantitatively assess the performance of models, including peak signal-to-noise ratio (PSNR), structural similarity index (SSIM), and normalized mean squared error (NMSE).

### 3.1. Ablation Studie

To verify the effectiveness of each key module, we conduct a series of ablation studies by (a) using CPNet as the backbone for SPET coarse prediction, (b) initializing CPNet encoder with PretrainNet pre-trained with only self-reconstruction task, (c) initializing CPNet encoder with PretrainNet pre-trained with both self-reconstruction and dose classification tasks, and (d) further introducing the RefineNet for refinement. The quantitative results are shown in Table 1. By comparing the results of (a), (b) and (c), we find that our pre-training phase can effectively enhance the feature extraction ability of the network, improving the PSNR at three DRFs by 0.799dB, 0.936dB, and 1.465dB, respectively. Notably, the pre-trained PretrainNet attains an accuracy of 79.6% on the tracer dose classification task, indicating that PretrainNet can extract discriminative features. Compared with (c), it can be found that the coarse-to-fine design in our method boosts the PSNR by 0.156dB, 0.768dB, and 2.413dB for DRF values of 100, 50 and 20, respectively. Thus, we can draw the conclusion that all the key modules contribute to boosting the RPET image quality.

### 3.2. Comparison with State-of-the-Art (SOTA) Methods

To demonstrate the superiority of our model, we compare our method with several SOTA PET reconstruction methods, containing Auto-context [2], StackGAN [3], Ea-GAN [11], AR-GAN [5], modified cGAN [7], LH-FrequencyNet [12]. All the comparative methods utilize mixed LPET images of multiple DRFs as input, and the quantitative results are reported in Table 2. Moreover, to verify the validity of our model for reconstructing multi-dose-level LPET images, we separately train the U-Net [9] at each DRF as the benchmark U-Net (Separate) for comparison, and the results are shown in the second row in Table 2. As can be observed, all comparison methods promote boosting LPET image quality improvement. Yet, it should be noted that Auto-context, StackGAN and Ea-GAN exhibit worse synthesis results in all cases when compared with U-Net (Separate). We reason that this may be attributed to the separate training of the U-Net for LPET images of each DRF, as the U-Net (Separate) only needs to learn the mapping between single-dose-level LPET and SPET, while other methods are forced to capture the complex mapping relationship between LPET images at multiple dose levels and SPET images. In contrast, our method consistently outperforms the comparison methods at all DRFs. Compared with U-Net (Separate), our method increases 0.579dB PSNR and 0.006 SSIM, and decreases 0.0432 NMSE for DRF =100, improving the PSNR by 0.373dB and 0.336dB for DRF = 50 and DRF = 20, respectively, suggesting the validity of our method in optimizing LPET images at multiple DRFs. Meanwhile, the paired t-test is also performed to check whether the improvements are statistically significant. As indicated by '*', the p-values are smaller than 0.05 in most cases, showing that our method is significantly better than the compared methods.

We also present the visual comparison results in Fig. 2. It can be observed that the RPET images reconstructed by our method are closest to the ground truth and present sharper textures and more details compared with other approaches. The error maps also show that our method obtains the RPET images most resembling the target SPET images, as evidenced by the darkest color of error maps in all three DRF cases, which means that the difference between the RPET images and the target SPET images is minimal. In general, both qualitative and quantitative results demonstrate that our method successfully enables multi-dose-level PET reconstruction with a single model.

**Table 1.** Ablation study results on Ultra-low Dose PET Imaging Challenge 2022 dataset

| Variant Model | DRF=100 | | | DRF=50 | | | DRF=20 | | |
|---|---|---|---|---|---|---|---|---|---|
| | PSNR | SSIM | NMSE | PSNR | SSIM | NMSE | PSNR | SSIM | NMSE |
| (a) | 29.232 | 0.953 | 0.0395 | 30.258 | 0.963 | 0.0283 | 30.314 | 0.968 | 0.0266 |
| (b) | 29.341 | 0.957 | 0.0382 | 30.299 | 0.968 | 0.0272 | 30.909 | 0.974 | 0.0254 |
| (c) | 30.031 | 0.964 | 0.0314 | 31.194 | 0.970 | 0.0234 | 31.779 | 0.976 | 0.0200 |
| **(d)** | **30.187** | **0.960** | **0.0311** | **31.962** | **0.973** | **0.0223** | **34.192** | **0.982** | **0.0121** |

**Table 2.** Comparative experimental results on Ultra-low Dose PET Imaging Challenge 2022 dataset

| Variant Model | DRF=100 | | | DRF=50 | | | DRF=20 | | |
|---|---|---|---|---|---|---|---|---|---|
| | PSNR | SSIM | NMSE | PSNR | SSIM | NMSE | PSNR | SSIM | NMSE |
| LPET | 22.100* | 0.893* | 1.878* | 25.713* | 0.941* | 0.288* | 30.199* | 0.965* | 0.0418* |
| U-Net (Separate) | 29.608 | 0.954* | 0.0743 | 31.589 | 0.971 | 0.0349* | 33.856 | 0.982 | 0.0178* |
| Auto-context | 27.354* | 0.958 | 0.0487* | 28.158* | 0.966* | 0.0408* | 28.685* | 0.972* | 0.0364* |
| StackGAN | 28.778* | 0.958 | 0.0407* | 29.913* | 0.968* | 0.0308* | 30.801* | 0.977* | 0.0251* |
| Ea-GAN | 29.278* | 0.961 | 0.0383* | 30.511* | 0.969 | 0.0291* | 31.596* | 0.977* | 0.0229* |
| AR-GAN | 29.448* | 0.959 | 0.0392* | 30.770* | 0.970 | 0.0277* | 32.067* | 0.978 | 0.0207* |
| modified cGAN | 29.747 | 0.961 | 0.0429 | 31.111 | 0.972 | 0.0292* | 32.841* | 0.981 | 0.0189* |
| LH-FrequencyNet | 29.938 | 0.963 | 0.0427 | 31.425 | 0.972 | 0.0273 | 33.302* | 0.982 | 0.0169* |
| **Proposed** | **30.187** | **0.960** | **0.0311** | **31.962** | **0.973** | **0.0223** | **34.192** | **0.982** | **0.0121** |

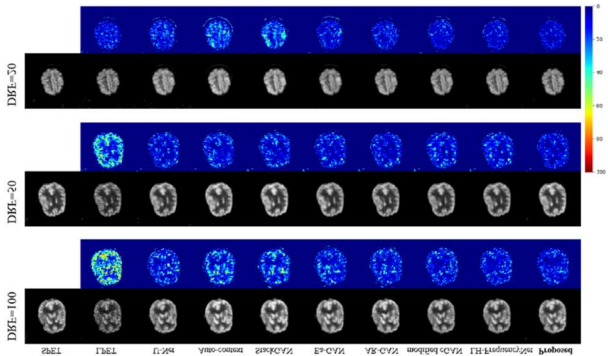

Fig. 2 Visual comparison results on MICCAI 2022 Ultra-low Dose PET Imaging Challenge Dataset at DRF = 100, 50, 20.

## 4. CONCLUSION

In this paper, to alleviate the limitation of existing single-dose-level LPET reconstruction methods, we propose a novel two-phase multi-dose-level PET image reconstruction method, including a pre-training phase and a SPET prediction phase. The pre-training phase incorporates dose classification and reconstruction tasks to respectively exploit fine-grained discriminative features and effective semantic representation in multi-dose-level LPET images, thus enhancing the feature extraction capability of the model. To generate RPET images more closely resemble the real SPET images the SPET prediction phase adopts a coarse prediction network and a refinement network, which gradually reduce the differences between LPET at different doses and target SPET images. Extensive experiments conducted on a public dataset demonstrate the generality and superiority of our method, compared with state-of-the-art reconstruction approaches.

## 5. COMPLIANCE WITH ETHICAL STANDARDS

This study was conducted retrospectively using human subject data made available in open access. Ethical approval was *not* required as confirmed by the license attached with the open access data. This article does not contain any studies with animals performed by any of the authors.

## 6. ACKNOWLEDGMENTS

This work is supported by the National Natural Science Foundation of China (NSFC 6237132, 62071314), Sichuan Science and Technology Program 2023YFG0263, 2023YFG0025, 2023NSFSC0497.

## REFERENCES

[1] A. B. de González and S. Darby, "Risk of cancer from diagnostic X-rays: estimates for the UK and 14 other countries.," *The Lancet*, vol. 363, no. 9406, pp. 345–351, Jan. 2004.

[2] L. Xiang, Y. Qiao, D. Nie, L. An, W. Lin, Q. Wang, and D. Shen, "Deep auto-context convolutional neural networks for standard-dose PET image estimation from low-dose PET/MRI," *Neurocomputing*, vol. 267, pp. 406–416, Dec. 2017.

[3] Y. Wang, B. Yu, L. Wang, C. Zu, D. S. Lalush, W. Lin, X. Wu, J. Zhou, D. Shen, and L. Zhou, "3D conditional generative adversarial networks for high-quality PET image estimation at low dose," *NeuroImage*, vol. 174, pp. 550–562, Jul. 2018.

[4] Y. Lei, X. Dong, T. Wang, K. Higgins, T. Liu, W. J. Curran, H. Mao, J. A. Nye, and X. Yang, "Whole-body PET estimation from low count statistics using cycle-consistent generative adversarial networks," *Physics in Medicine and Biology*, vol. 64, no. 21, pp. 215017, Nov. 2019.

[5] Y. Luo, L. Zhou, B. Zhan, Y. Fei, J. Zhou, Y. Wang, and D. Shen, "Adaptive Rectification based Adversarial Network with Spectrum Constraint for High-quality PET Image Synthesis," *Medical Image Analysis*, vol. 77, pp. 102335, Apr. 2022.

[6] S. Xue, R. Guo, K. P. Bohn, J. Matzke, M. Viscione, I. Alberts, H. Meng, C. Sun, M. Zhang, M. Zhang, R. Sznitman, G. El Fakhri, A. Rominger, B. Li, and K. Shi, "A cross-scanner and cross-tracer deep learning method for the recovery of standard-dose imaging quality from low-dose PET.," European Journal of Nuclear Medicine and Molecular Imaging, vol. 49, no. 6, pp. 1843–1856, May 2022.

[7] L. Zhang, Z. Xiao, C. Zhou, J. Yuan, Q. He, Y. Yang, X. Liu, D. Liang, H. Zheng, W. Fan, X. Zhang, and Z. Hu, "Spatial adaptive and transformer fusion network (STFNet) for low-count pet blind denoising with MRI.," *Medical Physics*, vol. 49, no. 1, pp. 343–356, Jan. 2022.

[8] K. Zhang, W. Zuo, and L. Zhang, "FFDNet: Toward a Fast and Flexible Solution for CNN-Based Image Denoising," *IEEE Transactions on Image Processing*, vol. 27, no. 9, pp. 4608-4622, Sept. 2018.

[9] O. Ronneberger, P. Fischer, and T. Brox, "U-Net: Convolutional Networks for Biomedical Image Segmentation," in *Lecture Notes in Computer Science, Medical Image Computing and Computer-Assisted Intervention – MICCAI 2015*, Springer, 2015, vol. 9351, pp. 234-241.

[10] S. Chen, G. Bortsova, A.-U. Juarez, G. Tulder, and M. Bruijne, "Multi-Task Attention-Based Semi-Supervised Learning for Medical Image Segmentation," in *Lecture Notes in Computer Science, Medical Image Computing and Computer-Assisted Intervention – MICCAI 2019*, 2019, vol. 11766, pp. 457-465.

[11] B. Yu, L. Zhou, L. Wang, Y. Shi, J. Fripp, and P. Bourgeat, "Ea-GANs: Edge-Aware Generative Adversarial Networks for Cross-Modality MR Image Synthesis," *IEEE Transactions on Medical Imaging*, vol. 38, no. 7, pp. 1750-1762, Jul. 2019.

[12] C. Jiang, Y. Pan, Z. Cui, and D. Shen, "Reconstruction of Standard-Dose Pet From Low-Dose Pet Via Dual-Frequency Supervision and Global Aggregation Module," in *2022 IEEE 19th International Symposium on Biomedical Imaging (ISBI)*, Kolkata, India, 2022, pp. 1-5.